\documentclass{elsart}
\usepackage{epsfig}
\usepackage{amssymb}

\def\hc#1{\hbox to\hsize{\hss #1\hss}}

\journal{{\tt Astroparticle Physics}}

\runauthor{T.~Antoni et al. (KASCADE Collaboration) }
\runtitle{Muon Density Measurements}

\begin{document}

\begin{frontmatter}

\title{Muon Density Measurements with the KASCADE Central Detector}

\author[KA-FZK]{T.~Antoni},
\author[KA-FZK]{W.\,D.~Apel},
\author[BU]{F.~Badea},
\author[KA-FZK]{K.~Bekk},
\author[KA-FZK]{K.~Bernl\"ohr\thanksref{nowatBerlin}},\relax 
   \thanks[nowatBerlin]{now at: Humboldt University Berlin, Germany}
\author[KA-FZK,KA-Uni]{H.~Bl\"umer},
\author[KA-FZK]{E.~Bollmann},
\author[BU]{H.~Bozdog},
\author[BU]{I.\,M.~Brancus},
\author[KA-FZK]{C.~B\"uttner},
\author[YE]{A.~Chilingarian},
\author[KA-Uni]{K.~Daumiller},
\author[KA-FZK]{P.~Doll},
\author[KA-FZK]{J.~Engler},
\author[KA-FZK]{F.~Fe{\ss}ler},
\author[KA-FZK]{H.\,J.~Gils},
\author[KA-Uni]{R.~Glasstetter},
\author[KA-FZK]{R.~Haeusler},
\author[KA-FZK]{A.~Haungs\thanksref{corres}},\relax 
   \thanks[corres]{corresponding author; email: haungs@ik3.fzk.de}
\author[KA-FZK]{D.~Heck},
\author[KA-FZK]{T.~Holst},
\author[KA-Uni]{J.\,R.~H\"orandel},
\author[KA-FZK,KA-Uni]{K.-H.~Kampert},
\author[LZ-Dep]{J.~Kempa\thanksref{nowatWarsaw}},\relax 
   \thanks[nowatWarsaw]{now at: Warsaw University of Technology, Poland}
\author[KA-FZK]{H.\,O.~Klages},
\author[KA-Uni]{J.~Knapp\thanksref{nowatLeeds}},\relax 
   \thanks[nowatLeeds]{now at: University of Leeds, U.K.}%
\author[KA-FZK]{K.\,U.~K\"ohler\thanksref{nowatPSI}},\relax 
   \thanks[nowatPSI]{now at: ETH Zurich, Switzerland}%
\author[KA-FZK]{G.~Maier},
\author[KA-FZK]{H.-J.~Mathes},
\author[KA-FZK]{H.\,J.~Mayer},
\author[KA-FZK]{J.~Milke},
\author[KA-FZK]{M.~M\"uller},
\author[KA-FZK]{J.~Oehlschl\"ager},
\author[BU]{M.~Petcu},
\author[KA-FZK]{H.~Rebel},
\author[KA-FZK]{M.~Risse},
\author[KA-FZK]{M.~Roth},
\author[KA-FZK]{G.~Schatz\thanksref{nowaHabicht}},\relax 
   \thanks[nowaHabicht]{present address: Habichtweg 4, D-76646 Bruchsal, Germany}
\author[KA-FZK]{J.~Scholz},
\author[YE]{S.\,H.~Sokhoyan},
\author[KA-FZK]{T.~Thouw},
\author[KA-FZK]{H.~Ulrich},
\author[BU]{B.~Vulpescu\thanksref{nowatHD}},\relax 
   \thanks[nowatHD]{now at: University of Heidelberg, Germany}
\author[KA-Uni]{J.\,H.~Weber},
\author[KA-FZK]{J.~Wentz},
\author[KA-FZK]{J.~Wochele},
\author[LZ-Sol]{J.~Zabierowski},
\author[KA-FZK]{S.~Zagromski} 

\collab{(The KASCADE Collaboration)}

\address[KA-FZK]{Institut f\"ur Kernphysik, Forschungszentrum Karlsruhe,
      	     76021~Karlsruhe, Germany}
\address[BU]{National Institute of Physics and Nuclear Engineering, 
             7690~Bucharest, Romania}
\address[YE]{Cosmic Ray Division, Yerevan Physics Institute, 
             Yerevan~36, Armenia}
\address[KA-Uni]{Institut f\"ur Experimentelle Kernphysik, University of
             Karlsruhe, 76021~Karlsruhe, Germany}
\address[LZ-Dep]{Department of Experimental Physics, 
             University of Lodz, 90236~Lodz, Poland}
\address[LZ-Sol]{Soltan Institute for Nuclear Studies,
             90950~Lodz, Poland}
	     
\ifx AA
\makeatletter
\begingroup
  \global\newcount\c@sv@footnote
  \global\c@sv@footnote=\c@footnote     
  \output@glob@notes  
  \global\c@footnote=\c@sv@footnote     
  \global\t@glob@notes={}
\endgroup
\makeatother
\fi

\newpage

\begin{abstract}
\noindent Frequency distributions of local muon densities in 
high-energy extensive air-showers (EAS) are presented as 
signature of the primary cosmic ray energy spectrum in 
the knee region. 
Together with the gross shower variables like shower core 
position, angle of incidence, and the shower sizes, 
the KASCADE experiment is able to measure local muon densities 
for two different muon energy thresholds. 
The spectra have been reconstructed 
for various core distances, as well as for particular subsamples, 
classified on the basis of the shower size ratio $N_\mu/N_{\rm e}$. 
The measured density spectra of the total sample exhibit clear 
kinks reflecting the knee of the primary energy spectrum. 
While relatively sharp changes of the slopes are observed in the  
spectrum of EAS with small values of the shower size ratio,
no such feature is detected at EAS of large $N_\mu/N_{\rm e}$
ratio in the energy range of 1--10 PeV.
Comparing the spectra for various thresholds and core distances
with detailed Monte Carlo simulations the validity of EAS simulations
is discussed.

\end{abstract}


\end{frontmatter}


\section {Introduction}
\label{sec:intro}

Measurements of the energy spectrum and the elemental composition 
of the primary cosmic radiation constrain 
theoretical models of the sources, acceleration mechanisms and 
transport of the radiation through the interstellar space.  
While for lower energies direct measurements by satellites or
balloon-borne detectors yield spectroscopic results 
(see ref.\cite{wiebel-1998}), for primary energies above some 
$10^{14}\,$eV only indirect measurements via extensive air shower 
(EAS) observations can be performed.
It is well known that the energy spectrum of the primary cosmic
radiation shows a kink (mostly referred to as ``knee'') at
energies around 3~PeV \cite{kalmy-1995}. 
Though the first evidence of the existence of this knee has been 
presented more than 40 years ago \cite{khrist-1958},
the knowledge of the detailed structure of the spectrum in the 
PeV region is still scarce, and the origin of the knee not yet 
understood. \\
Most of the earth-bound air shower experiments use large 
detector arrays to measure charged particles 
and reconstruct shower sizes of the individual 
events by adjusting a particular lateral distribution
function to the measured densities. 
The resulting shower size spectra 
reflect the primary energy spectrum, but a quantitative conversion 
to energy has to invoke a model of the shower development and on an 
assumption of a mass composition. 
Hence the determination of the energy spectrum is affected by 
different systematic uncertainties, especially by the  
dependence on the model of high-energy interactions. This also leads
to a mutual dependence of the 
results for the energy spectrum and mass composition. \\   
When comparing recent results of earth-bound air shower experiments, 
like CASA-MIA~\cite{casa-1999}, TIBET~\cite{tibet-1993} or 
Akeno~\cite{akeno-1984}, 
significant differences in the absolute magnitudes of the total flux, 
the knee position and slope of the energy spectrum are noticed.
A recent non-parametric analysis of KASCADE data \cite{antoni-non}
reports equally large differences on the energy spectra, 
depending on the high-energy hadronic interaction model, 
illustrating the considerable influence of the interaction 
models underlying the Monte Carlo simulations. 
To identify the ultimate sources of the disagreements, 
it would be useful to analyze different experiments on basis 
of a coherent methodology as well as to compare the resulting 
features for various sets of different EAS parameters in the
individual experiments.  \\
In the present paper we endeavor to analyze  
the frequency distribution of local muon densities at 
fixed distances from the shower core. 
The local muon density spectra reflect the gross features of the 
primary energy spectrum, as the muon content for a certain
distance to the shower center observed at sea-level 
is mainly determined by the primary energy.
While the reconstruction of electron or muon size spectra 
necessarily implies a choice of the form of the lateral 
distribution function, spectra of the muon density 
are free from this bias. 
Thus with ``independent'' measurements of such spectra 
for different fixed core distances allow a check on 
the lateral distribution obtained from simulations. 
In addition, the layout of the KASCADE experiment~\cite{Klages-1997}, 
with a central detector system consisting of densely 
packed muon counters with different shielding, enables 
the study of density spectra for two different muon 
energy thresholds. 
Hence the consistency of the simulations with respect to the 
muon energy spectrum can be performed.


\section {Experimental setup and data handling}
\label{exper}

\begin{figure}[ht]
\hc{\epsfig{file=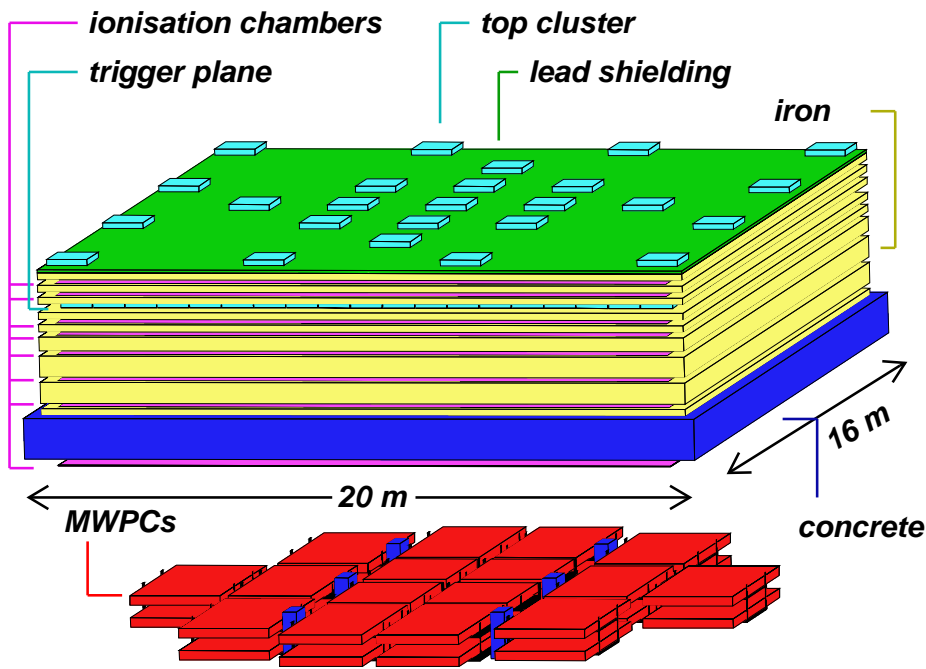,clip=,width=\hsize}}
\caption[KASCADE Central Detector]{Schematic view of the 
KASCADE central detector. It consists
of four detector systems: an 8-layer hadron 
calorimeter~\cite{Engler-1999}, scintillation counters at the
trigger plane and top cluster, and a setup of multiwire proportional
chambers (MWPC)~\cite{MWPC-2000} below the calorimeter.}
\label{fig:calo}
\end{figure}

KASCADE (KArlsruhe Shower Core and Array DEtector)
is a multi-detector setup \cite{Klages-1997}
at Forschungszentrum Karlsruhe ($110\,$m a.s.l.), 
Germany, for EAS measurements 
in the primary energy range around the knee. 
The main detector components of KASCADE used for the 
present analysis are an
``array'' of 252 stations, located on a squared grid
with $13\,$m spacing and a ``central detector'' comprising
additional detector systems. \\
The array is organized in 16 subarrays ($4 \times 4\,$ 
stations each) and provides the data necessary for the 
reconstruction of the basic EAS characteristics like  
electron and muon size (total number of electrons and muons 
in the EAS), core location, and arrival direction of individual
air showers. 
The special arrangement of shielded and unshielded detectors 
on top of each other allows an independent
estimation of the total electron and muon number
for each individual shower. The densities are estimated
and corrected iteratively for punch-through effects ($\mu$-counters)
and muon contamination ($e/\gamma$-counters).
Lateral correction functions from simulations of EAS and 
detectors are used.
The reconstructed particle densities are fitted by
Nishimura-Kamata-Greisen (NKG) functions in the experimental 
accessible distance ranges (10-200$\,$m for the
electron component, 40-200$\,$m for the muon component).
The densities obtained are then integrated from zero to infinity for
the total numbers $N_e$ and $N_\mu$. In addition we quote 
the so-called truncated numbers for which the NKG functions 
are integrated in a limited range only
($N_e^{\rm tr}=\int_{10m}^{120m}\rho_e \cdot 2 \pi R dR$ and
 $N_\mu^{\rm tr}=\int_{40m}^{200m}\rho_\mu \cdot 2 \pi R dR$). 
These truncated numbers provide reduced systematic uncertainties
since extrapolations into the radial range outside our 
measurement areas are avoided.
Uncertainties are estimated by Monte Carlo calculations and range
below 20\% for the total numbers and even better for the 
truncated numbers.
The location of the shower core once inside a fiducial area
is determined to better than 3$\,$m. 
The arrival direction of the shower is reconstructed from the 
arrival times of EAS particles ($\sigma \approx 0.5^\circ$).
These procedures are described in detail elsewhere~\cite{antoni-lat}. \\
The KASCADE central detector is placed at the geometrical 
center of the detector array. It consists
of four different detector systems (Fig.~\ref{fig:calo}),
covering a total area of 16$\times$20$\,$m$^2$.
The local muon density of EAS is measured
with the multiwire proportional chambers (MWPC)
and the trigger plane. \\
Below the hadron calorimeter~\cite{Engler-1999}, 
with a total thickness of
5$\,$cm lead, 154$\,$cm steel, and 77$\,$cm concrete corresponding
to a threshold for vertical muons of 2.4$\,$GeV, 
a setup of 32 large 
multiwire proportional chambers is installed 
\cite{MWPC-2000}. 
The chambers of three different sizes
(5$\,$m$^2$, 8.2$\,$m$^2$, and 8.8$\,$m$^2$)
are arranged in two layers with 38$\,$cm vertical distance.
In total each layer has a sensitive area of 129$\,$m$^2$. 
A single chamber consists of two layers of cathode strips
at angles of $\pm34^\circ$ with respect to a layer of 
anode wires in between.
The chambers are operated with an argon-methane gas mixture. 
The electronics allows a digital readout of all wires and strips,
i.e. in total of c. 32,000 channels. 
Hits of through-going particles are reconstructed as the  
intersection of the anode wires and the cathode strips.  
Thresholds and delays are adjustable for each channel separately. 
A continuous monitoring of the reconstruction efficiency
during measurements is performed.  
The chambers have a spatial resolution of about 5$\,$mm. \\
The reconstruction of high-energy muons in the MWPC
starts from the reconstructed hits in each plane 
and the shower direction. The reconstructed direction is 
required to agree with the shower direction within 
$\pm 15^\circ$ in zenith and $\pm 45^\circ$ in azimuth
(the azimuth cut is not used for showers 
with zenith angles of $<10^\circ$).
These cuts appear reasonable as for core distances below 100$\,$m 
high-energy muon tracks are nearly parallel to the shower axis.
It is known from simulations that in the considered range of primary 
energy (PeV) and core distances the muon
density (for $E_\mu > 2.4\,$GeV) very rarely exceeds 1 per m$^2$. 
Therefore reconstruction ambiguities are negligible.
High-energy $\delta$-electrons which are produced to
a small amount in the absorber are eliminated by
calculating the height of the intersection of two nearby tracks. 
If they cross inside the central detector, 
the track with the larger deviation from the shower
axis is rejected while the other is accepted as a muon only. 
About $0.1\%$ of tracks are rejected by this cut.  
The spatial resolution of single tracks is about 1.0$\,$cm,
the angular resolution is $\approx\,1^\circ$.
The number of tracked muons $N_\mu^\star$ is also corrected for 
the reconstruction efficiency which is estimated for each single 
data acquisition run ($\approx 12\,$h) separately.
The efficiency was found to be very stable with a mean value of 
$\langle \epsilon \rangle = 93\%$~\cite{MWPC-2000}.
The local muon density $\rho_\mu^\star$ for each EAS
is defined by $N_\mu^\star$ divided by the total sensitive 
area $A^\star$ of the MWPC setup.
Due to the layout of the chambers, $A^\star$ depends on the 
angle of incidence of the shower and is calculated for each event
individually ($\langle A^\star \rangle = 107\,$m$^2$ for the 
selected EAS). Only that area where muons parallel to the
shower axis would penetrate the whole absorber and both chamber 
planes, is taken into account for the calculation of the 
muon density. \\
The second detector system is a layer
of 456 plastic scintillation detectors in the third gap of the 
calorimeter, called trigger plane~\cite{Engler-1999}. 
Each detector consists of two square plates of 
plastic scintillators (47.5$\times$47.5$\times 3\,$cm$^3$)
separated by a wavelength-shifter, which is read out by a
single photomultiplier. Fast electronics records low-energy 
(muons) and high-energy deposits (cascading hadrons) and 
provides a trigger for the calorimeter and other detector systems.
In the present analysis the trigger plane with an active area 
of 208$\,$m$^2$ is used to estimate the local density 
of muons with a threshold of 490$\,$MeV for vertical incidence. \\
The muon density $\rho_\mu^{\rm tp}$ at the trigger plane
is reconstructed in the following way:
To remove signals from cascading hadrons in the absorber an 
upper limit of the energy deposit of $30\,$MeV in each 
of the 456 scintillation counters is imposed. Detectors
with larger energy deposits and their immediate neighbours 
are not considered for further reconstruction. 
For the remaining detectors, the energy deposit 
and the sensitive area, both corrected for the 
shower direction, are summed up.
The number of reconstructed muons $N_\mu^{\rm tp}$
is then calculated by the sum of the energy deposits divided
by the mean energy deposit of a single muon in the shower.
According to Monte Carlo calculations this mean value 
depends slightly on the core distance $R_{\rm c}$
($\langle E_{\rm dep} \rangle = 7.6211 - 0.00495 
\cdot R_{\rm c}\,$ in MeV) and is corrected for.
$R_{\rm c}$ is the distance in meter of the core 
position to the center of the trigger plane (or MWPC)
projected to a plane perpendicular to the 
shower axis.
The density $\rho_\mu^{\rm tp}$ is obtained as ratio of
$N_\mu^{\rm tp}$ and the sensitive area of the 
trigger plane for each individual event 
($\langle A^{\rm tp} \rangle = 202\,$m$^2$). \\
The MWPC setup is triggered by 
trigger plane and top cluster, but not by the detector array.
But the array, trigger plane and the top cluster 
are triggering all other components.
The trigger plane fires if more than 7 detectors have signals
$>1/3\,$mip or if at least one detector has 
$E_{\rm dep} > 300\,$MeV (for single hadron detection).
The top cluster triggers if more than 8 (out of 50) detectors 
show a signal. An array trigger is activated if 
half of the stations of at least one subarray 
show an energy deposit ($>1/3\,$mip).
One of the central detector triggers in conjunction with 
the array trigger have to be active to initiate the 
event reconstruction. \\
After some general cuts (core position less than $91\,$m
from array center, $\Theta < 40^\circ$, 
$lg(N_\mu^{\rm tr}) > 4.745 - 0.212\,{\rm lg}(N_e)$), more than 
two million events have been used for the present analysis,
recorded in circa 282 days of measuring time. 
Measured fluxes have been corrected by $9\%$ for the dead-time
of the data acquisition system.


\section {Local muon density spectra}
\subsection {All-particle spectra}
\label{spectra}

\begin{figure}[ht]
\hc{\epsfig{file=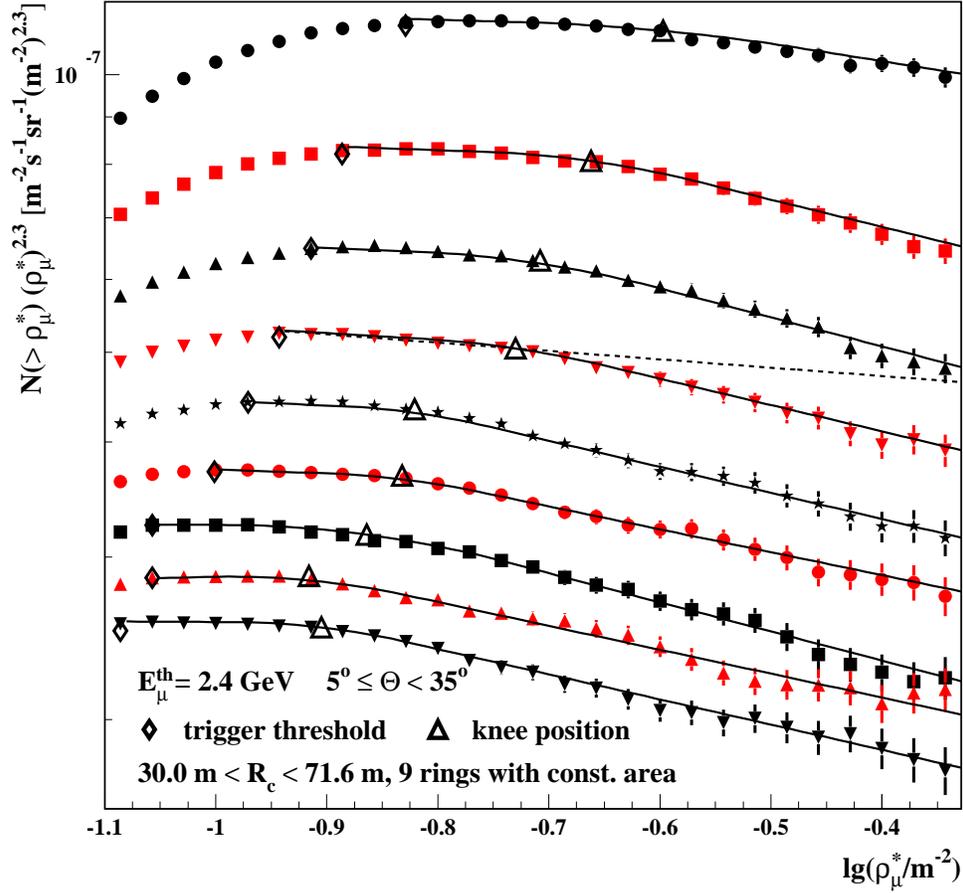,clip=,width=\hsize}}
\caption[Myon density spectra (MWPC)]{Integral spectra
of the local muon density $\rho_\mu^\star$ as measured by the 
MWPC system for different core distances $R_{\rm c}$ (from top 
to bottom with increasing $R_{\rm c}$). The upper limits of the 
radial bins are 37.0, 42.9, 48.0, 52.7, 57.0, 60.9, 64.7, 68.2, 
and 71.6$\,$m, respectively. The lines represent the results of 
the fit procedure (see text). The dashed line displays for one 
case the result of a fit with a single power law.}
\label{fig:specmw}
\end{figure}

\begin{figure}[ht]
\hc{\epsfig{file=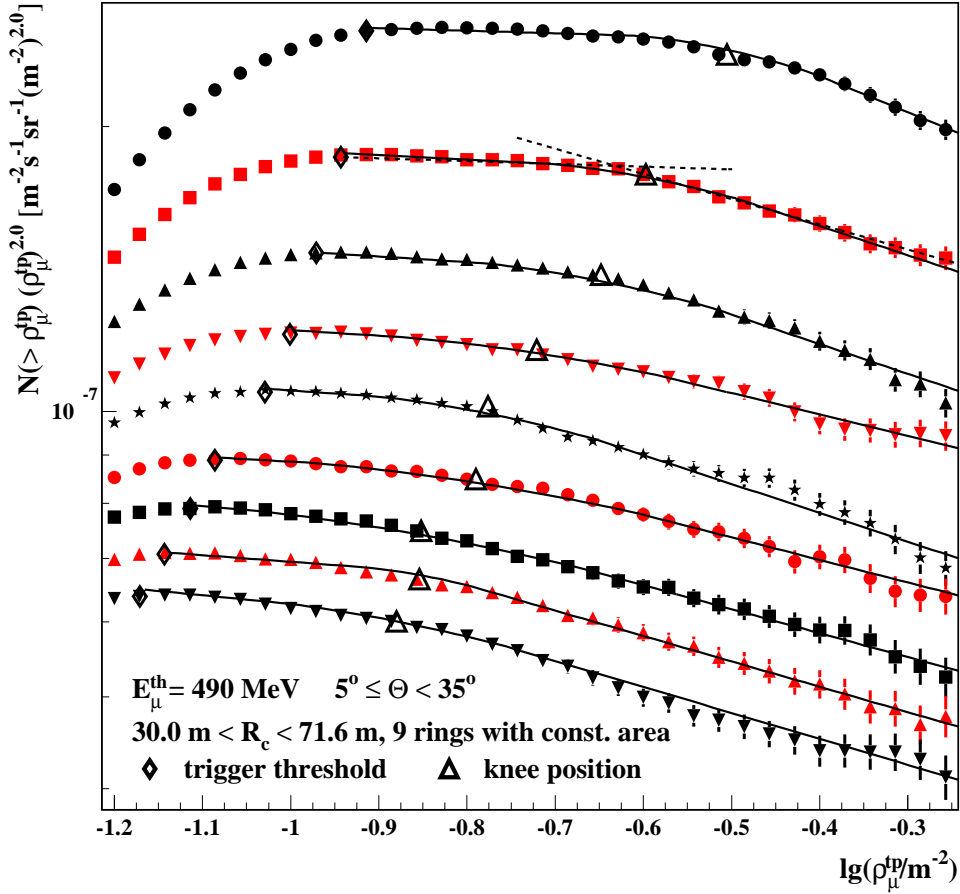,clip=,width=\hsize}}
\caption[Myon density spectra (trigger plane)]{Same as 
Figure~\ref{fig:specmw} but for the local muon density 
$\rho_\mu^{\rm tp}$ measured by the trigger plane.
Here the dashed lines display, for one case, the fit function 
of a two power law fit (below and above the knee position) 
without an intersecting knee region.}
\label{fig:spectp}
\end{figure}

It is reasonable to assume that at a fixed distance from the shower
axis the local muon densities map the energy of the primary 
particles~\cite{antoni-lat} and that muon density
spectra carry information about the primary energy spectra.  \\
The reconstruction of muon density spectra have been
performed for two energy thresholds and for nine 
core distance ranges (Figures~\ref{fig:specmw} and~\ref{fig:spectp}). 
These ranges are chosen in such a way that
the sampling area are of equal size (1473.4~m$^2$) and large enough 
to get reasonable statistical accuracy but retain small systematic 
uncertainties due to the extension of the core distance bins. 
To suppress punch-through effects of the hadronic or electromagnetic 
component, EAS with $R_{\rm c}<30\,$m are excluded.
EAS with $R_{\rm c}>72\,$m are excluded, too, because they
can have their core outside the KASCADE 
array ($R_0<91\,$m) if they are very inclined. 
The $\rho_\mu$-spectra are affected by trigger efficiencies
at low densities. The limitation at high densities are given 
by reconstruction uncertainties. 
For the MWPC system (higher energy threshold) these uncertainties  
begin at a fixed muon density of $\rho_\mu^{\star} \cong 
0.6\,$m$^{-2}$ due to ambiguities in the track reconstruction and
punch-through effects of cascading hadrons. The hadronic
energy and hadron particle density in EAS are increasing 
similarly to the muon density at all core distances~\cite{antoni-lat}. 
In case of the trigger plane the cut on the deposited energy  
($30\,$MeV) in each scintillation detector has a systematic 
influence on local densities above 
$\rho_\mu^{\rm tp} \cong 0.8\,$m$^{-2}$.
Here the density will be reduced since the intrinsic density 
fluctuations in the EAS together with fluctuations in the energy 
deposit lead to detector signals exceeding $30\,$MeV without 
punch-through effects. \\
Figure~\ref{fig:specmw} and Figure~\ref{fig:spectp} show the 
flux spectra for the two muon thresholds in integral form. 
The flux values are multiplied by 
$(\rho_\mu^\star)^{2.3}$ and $(\rho_\mu^{\rm tp})^{2.0}$,
respectively.
All spectra show a slight, but significant kink with decreasing
density for increasing core distance. 
For the fit procedure the flux $lg(\frac{dN}{d \rho_\mu})$ 
is assumed to follow a power law below and above the knee region.
The following form of the differential spectra is assumed:
\begin{displaymath}
 lg(\frac{dN}{d \rho_\mu})=
\left\{ \begin{array}{ccc}
{b_{1}+\beta_{1}lg(\rho_\mu)} & 
\hspace*{0.4cm }{\rm for} \hspace*{0.2cm } &
   lg({\rho_\mu^{(1)}})\leq{lg(\rho_\mu})\leq{lg(\rho_\mu^{(2)})}\\
{a[b-lg(\rho_\mu)]^3+c} &
\hspace*{0.4cm } {\rm for} \hspace*{0.2cm } &
   lg({\rho_\mu^{(2)})<lg(\rho_\mu)<lg(\rho_\mu^{(3)})}\\
{b_{2}+\beta_{2}lg(\rho_\mu)} & 
\hspace*{0.4cm }{\rm for}\hspace*{0.2cm } &
   lg({\rho_\mu^{(3)})}\leq{lg(\rho_\mu)}\leq{lg(\rho_\mu^{(4)})}

\end{array} \right .
\end{displaymath}
The fit procedure estimates the indices $\beta_i$ of 
these power laws, the position of the knee (if existing),
and the boundaries of the different regions.
The values of the boundaries are estimated by the method of 
finite and dividing differences \cite{soko-1999}. 
Especially the lower thresholds of the spectra, where they begin 
to deviate from a power law dependence are defined by this method. 
With increasing core distances these trigger thresholds move to 
lower muon densities because of the decreasing lateral distribution
function.
The position of the knee is calculated as the weighted center of 
gravity of the bins inside the knee region.
The fit functions are included in
Figures~\ref{fig:specmw} and \ref{fig:spectp}, as well as
the particular position of the bending.
\begin{figure}[ht]
\hc{\epsfig{file=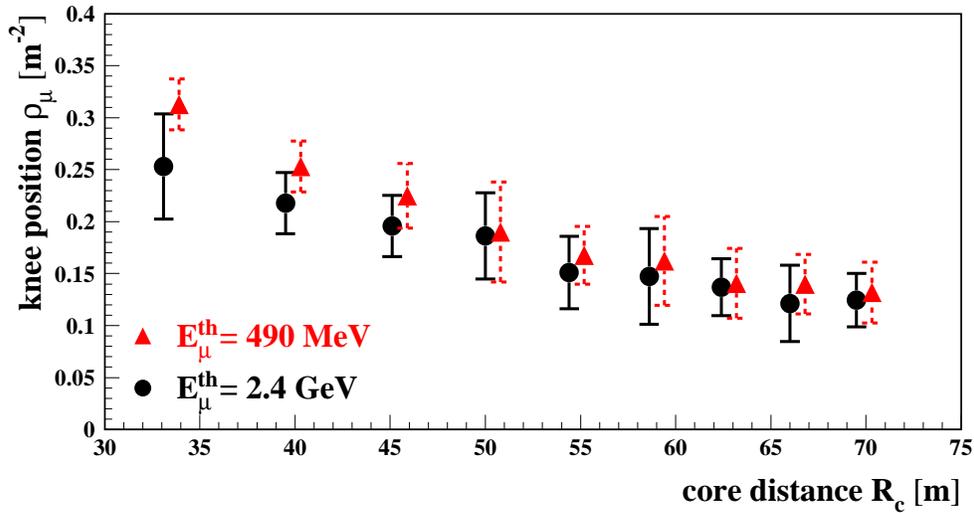,clip=,width=\hsize}}
\caption[Knee positions of muon spectra]{Knee positions of the 
muon density spectra vs. core distance for both energy thresholds. 
The error bars indicate the uncertainties of the fit 
procedure.}
\label{fig:knee}
\end{figure}
\begin{figure}[ht]
\hc{\epsfig{file=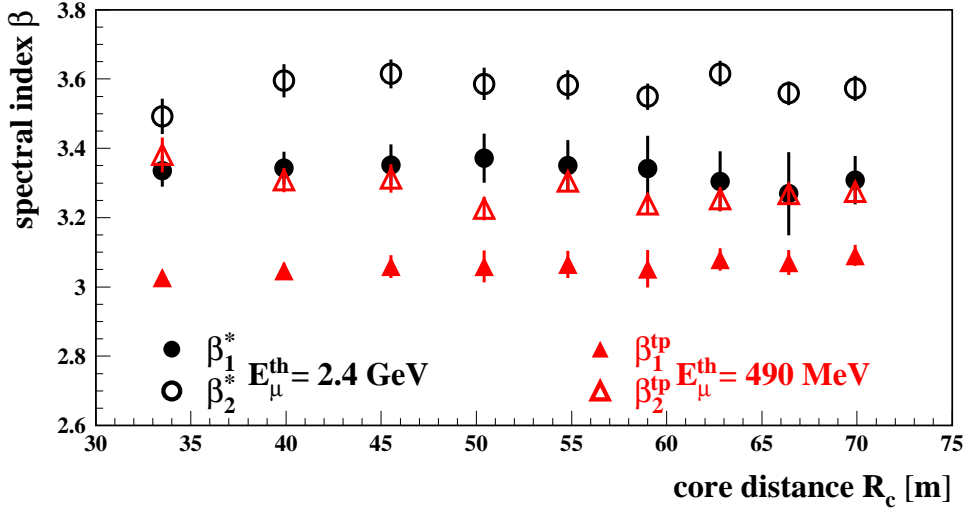,clip=,width=\hsize}}
\caption[Indices of muon spectra]{Power law indices of the 
differential muon density spectra for both energy thresholds. 
The error bars indicate the uncertainties of the fit 
procedure.}
\label{fig:betas}
\vspace*{0.5cm}
\end{figure}
The ``width'' of the knee region for all spectra amounts to 
$\Delta lg(\rho_\mu/{\rm m}^{-2}) \approx 0.15\,$. \\
It is a remarkable observation that within the statistical 
uncertainty the knee positions for all spectra
occur at the same EAS flux, for both thresholds
and all core distances (at a differential flux of 
$d{\rm N}/d\,lg({\rm \rho_\mu/ m}^{-2}) = (1.2 \pm 0.2)\cdot 10^{-6}\,$
m$^{-2}$s$^{-1}$sr$^{-1}\,$).
Figure~\ref{fig:knee} displays the knee positions 
for all reconstructed spectra. If we assume
that the knee is a feature of the primary energy spectrum,   
the data points mark the lateral distribution of muons for two
different energy thresholds for a fixed primary cosmic ray 
energy.
This shows that, for the EAS registered, the fraction of muons
between $490\,$MeV and $2.4\,$GeV is small.
Figure~\ref{fig:betas} shows the power law indices
of the density spectra. 
A higher muon energy threshold results in steeper 
spectra. This indicates a comparatively larger increase of the 
muon density per primary energy interval with increasing muon 
energy threshold.
The spectra for different core distances are almost parallel
leading to nearly constant indices for a given muon energy 
threshold. This confirms previous experimental 
results~\cite{antoni-lat} of only slight changes of the shape 
of the muon lateral distributions with increasing 
primary energy (which is different for the 
electromagnetic component of EAS). 
For both energy thresholds there is a clear difference
in the indices below and above the knee. \\ 
Several tests were performed to check the robustness of the
shape of the reconstructed spectra against
experimental uncertainties or systematic features of the analysis.
Differently chosen functional forms for the fit procedure to the 
differential spectra result in similar indices of the power laws. 
The significance of the knee remains stable (see also dashed 
lines in Figure~\ref{fig:spectp}). 
The assumption of a single power law for the spectra
(e.g. dashed line Figure~\ref{fig:specmw}) 
leads to reduced $\chi^2$-values
of around $1.5-2.$, whereas the used procedures fit
the data with $\chi^2$-values close to one. 
In addition, it has been found that effects of the 
binning of the density and of the core ranges, 
or of the azimuthal distribution of the showers are negligible 
and within the statistical uncertainty of the spectra.
Variations of the chosen zenith angle range shift the total 
spectrum in $\rho_\mu$ slightly, but the form of the spectra 
remains stable. This is reasonable, because differently
inclined showers (of the same primary energy) generate slightly
different numbers of muons, while the variation of the density
with energy does not change.

\subsection {Spectra of EAS subsamples}
\label{sec:spectras}

The ratio of the muon to electron content of EAS is traditionally
considered as a mass-sensitive observable~\cite{Blake-1998,jenny},
since heavy ion induced EAS tend to have a large ratio due to 
the faster development of the electromagnetic component 
in the atmosphere. 
The shower sizes $N_e$ and $N_\mu^{\rm tr}$
have been reconstructed for each individual event 
from the data of the array stations and the EAS have been
classified according to the ratio and then
\begin{figure}[hb]
\hc{\epsfig{file=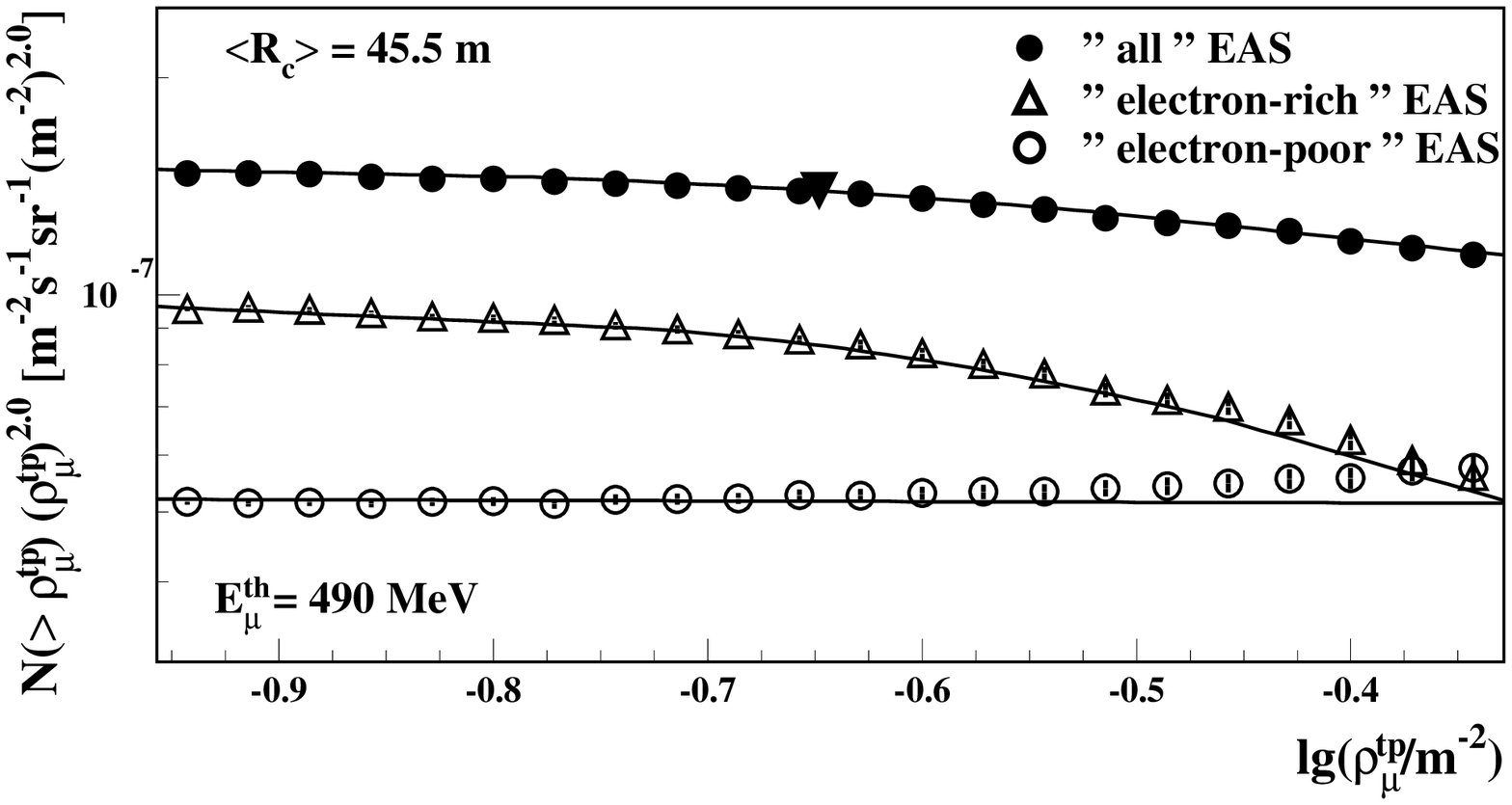,clip=,width=\hsize}}
\hc{\epsfig{file=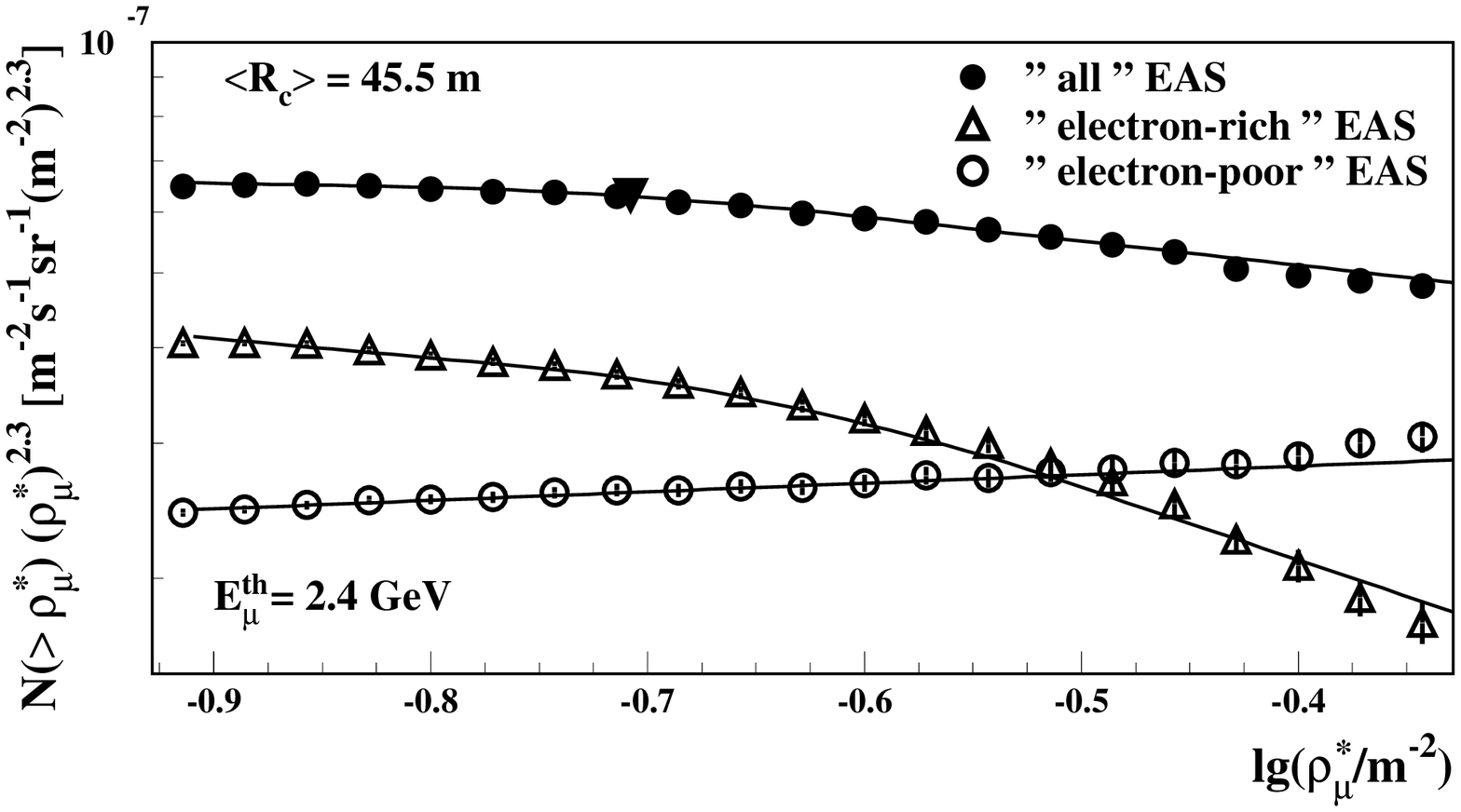,clip=,width=\hsize}}
\caption[Single mass group spectra]{Examples for measured 
spectra of different muon content. The ``all''-particle spectra have 
already been shown in Figures~\ref{fig:specmw}~and~\ref{fig:spectp} 
and are here compared with the spectra of ``electron-poor'' and 
``electron-rich'' EAS for the same core distance range.}
\label{fig:rspec}
\end{figure}
divided in ``electron-rich'' and ``electron-poor''.
The shower sizes are converted to the sizes of vertical showers
to eliminate the influence of the different zenith angles: \\
$ ln(N_e^\prime) = ln(N_e) - 
\frac{X}{\Lambda_e}\cdot(\sec{\theta} - 1) $ \\
$ln(N_\mu^\prime) = ln(N_\mu^{\rm tr}) - 
\frac{X}{\Lambda_\mu}\cdot(\sec{\theta} - 1) $ \\
where $N_e$, $N_\mu^{\rm tr}$, and $\theta$ are the reconstructed
quantities of the EAS, and $X = 1022\,$g/cm$^2$ is the 
observation level. 
The quantities $\Lambda_{e}$ and $\Lambda_{\mu}$ denote 
the absorption lengths of the electron and muon components 
in the atmosphere.
The values were obtained from 
Monte Carlo simulations and parameterised as
$\Lambda_e=104.3 + 13.5\cdot lg(N_e)\,$g/cm$^2$
and $\Lambda_\mu = 5\cdot \Lambda_e\,$. 
Especially the electron number depends significantly 
on the zenith angle due to the rapidly increasing atmospheric 
absorption.
The separation of the total sample of EAS in 
``electron-rich'' and ``electron-poor'' showers 
is performed by a cut in the ratio 
$Y_{\rm ratio} = lg(N_\mu^\prime) / lg(N_e^\prime) = 0.75\,$. 
This value is optimized by Monte Carlo calculations 
(see section~\ref{simul}).
The classification is done independently using
the local muon densities at the central detector. 
For both subsamples the spectra are 
deduced in the same way as the ``all-particle'' spectra. \\
As example Figure~\ref{fig:rspec} shows 
\begin{figure}[hb]
\hc{\epsfig{file=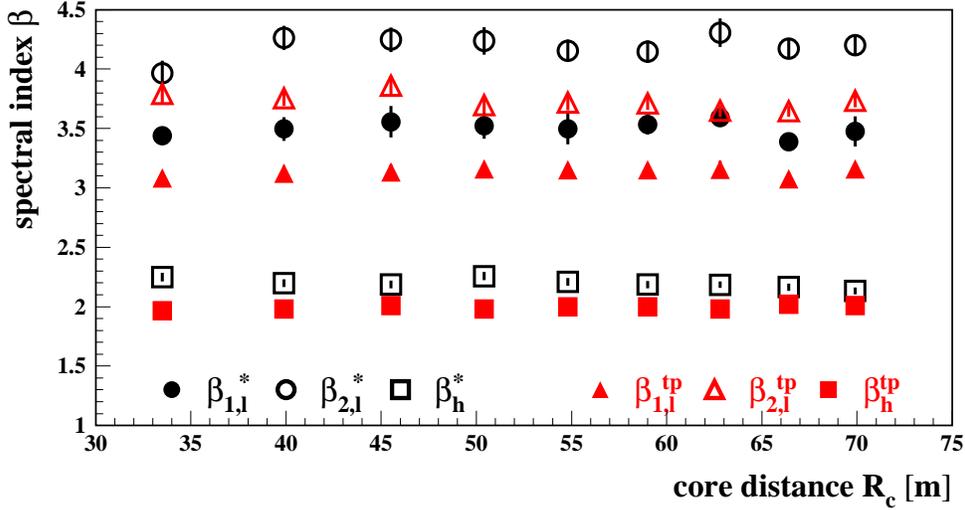,clip=,width=\hsize}}
\caption[Indices of mass group spectra]{Power law indices of the 
muon density spectra for both energy thresholds and all core distance
ranges for the samples of electron-rich ($\beta_l$ below and above 
the knee) and electron-poor ($\beta_h$) EAS. 
The error bars indicate the uncertainties of the fit procedure.}
\label{fig:betasr}
\end{figure}
the reconstructed local muon density spectra 
for $\langle R_{\rm c} \rangle = 45.5\,$m for all, for the 
electron-rich (predominantly light ion induced), and for the 
electron-poor (predominantly heavy ion induced) showers.
The general features of the spectra are similar for 
all core distance ranges; the component 
of electron-rich EAS dominates the 
flux below the knee while it strongly decreases after the kink.
No knee is seen in the component of electron-poor EAS, and the 
spectra can be described by a single power law. 
The resulting slopes of the spectra, especially the differences
of the slope-values for the two thresholds and subsamples,
are very similar for the various core distances as shown in
Figure~\ref{fig:betasr}. Whereas the assumed fit functions
describe the all-particle spectra well, the spectra for the 
electron-rich EAS are not well described by power laws
above the knee.
Also for the electron-poor sample slight deviations from a pure 
power law dependence at the high energy end are observed. 
This holds for all radial ranges.
An energy dependent separation efficiency of the primary masses as
well as astrophysical sources (composition, acceleration,
propagation) can cause these deviations from simple power
law dependencies.


\section {Comparisons with simulations}
\subsection {Air-shower simulations}
\label{simul}

For the interpretation of the measured muon density spectra
in terms of the primary energy spectrum a-priori knowledge 
inferred from Monte Carlo simulations of the air-shower 
development is necessary. 
The present analysis is based on CORSIKA simulations 
(version 5.62)~\cite{CORSIKA} and a full and detailed simulation 
of the detector response. The simulations have been performed
using the interaction model 
QGSJET~\cite{QGSJET} for the high-energy interactions and 
GHEISHA~\cite{GHEISHA} for interactions below 
$E_{\rm lab} = 80\,$GeV and subsequent decays.
The electromagnetic part of the showers is treated by 
EGS4~\cite{EGS}.
Observation level, earth's magnetic field, and the particle 
thresholds are chosen in accordance with the experimental 
situation of KASCADE. The U.S. standard atmosphere~\cite{CORSIKA}
was adopted.  
The simulations cover the energy range of $5\cdot10^{14}$ --
$3.06\cdot10^{16}\,$eV divided into 7 overlapping energy bins
with a spectral index of $-2.7\,$. For each bin 200 showers are 
simulated except for the two highest energy ranges where only 
100 and 50 showers were generated, respectively. 
The calculations are performed for three zenith angular ranges
($0^\circ-15^\circ$, $15^\circ-20^\circ$, $20^\circ-40^\circ$)
and for three primary masses: protons,  
oxygen and iron nuclei. 
The response of KASCADE is simulated by a detailed detector
simulation program based on the GEANT~\cite{GEANT} package.
Each generated shower is passed ten times through the detector
simulation.
The shower cores are randomly distributed over the KASCADE array 
within a circular area of $95\,$m radius around the center. 
Hence, a total statistics of 103,500 EAS is used. 
The output of the simulations is analyzed by the same procedures
\begin{figure}[ht]
\hc{\epsfig{file=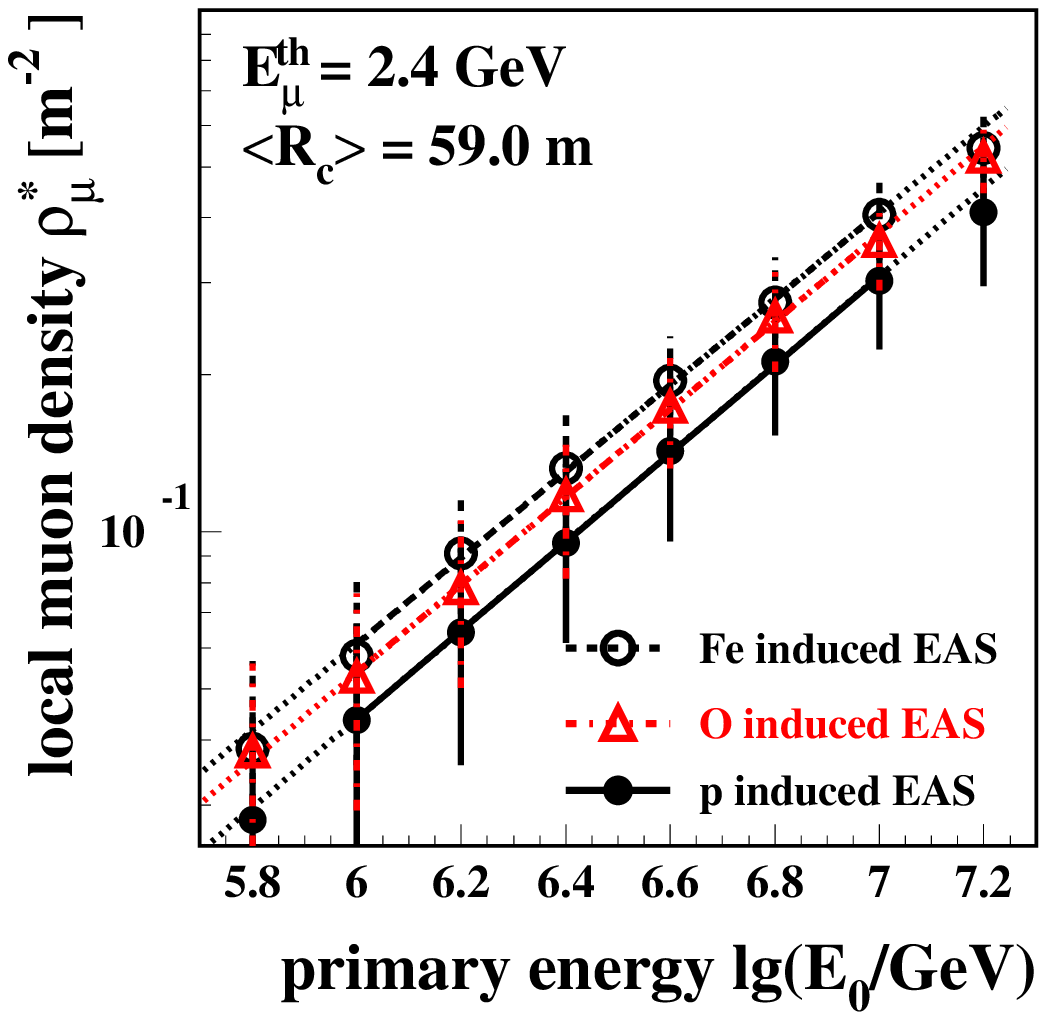,clip=,width=0.5\hsize}
    \epsfig{file=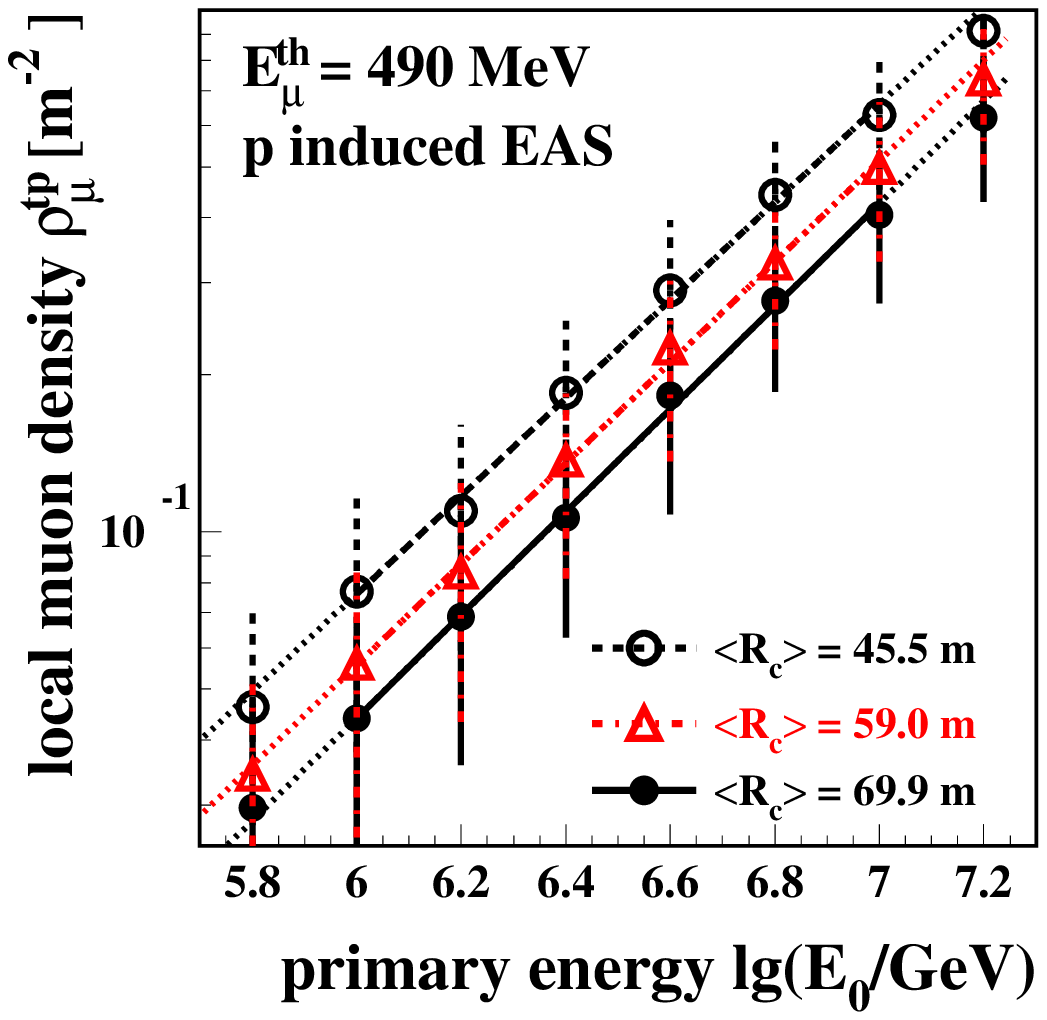,clip=,width=0.5\hsize}}
\caption[Muon density vs primary energy]{
Examples of local muon densities vs. primary energy of simulated 
EAS for different muon thresholds, 
core distances, and primary masses. The error bars indicate the
standard deviations of the densities. The lines show power law 
fits taking into account the statistical uncertainty
of the mean values which are smaller than the marker sizes.}
\label{fig:mcdens}
\end{figure}
\begin{figure}[ht]
\hc{\epsfig{file=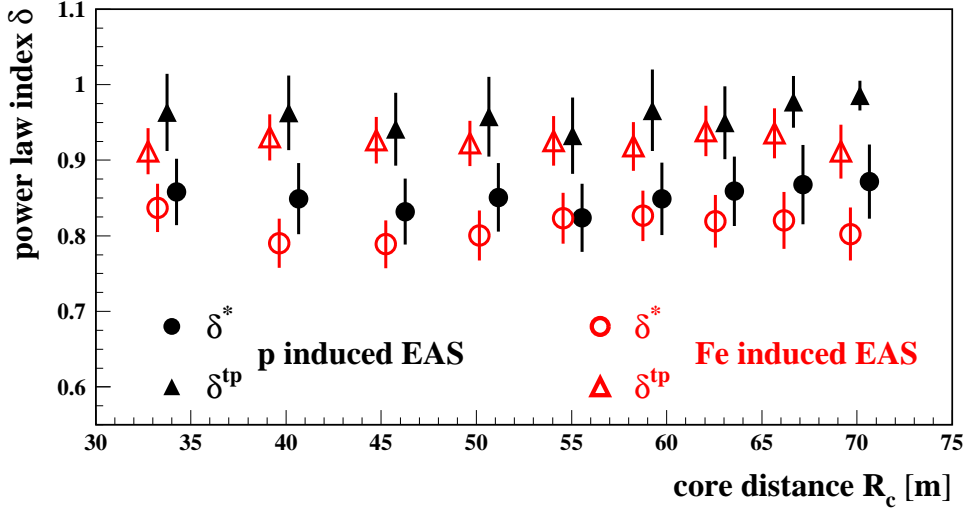,clip=,width=\hsize}}
\caption[Slopes of muon densities vs primary energy]{Power law
indices $\delta$ of the $\rho_\mu \propto E_0^\delta$ relation
for different $R_c$ and muon thresholds in case for primary 
protons and iron nuclei. 
The error bars indicate the systematic uncertainty
(see text).}
\label{fig:deltas}
\end{figure}
as applied to the measured data, reducing systematic uncertainties. \\
Figure~\ref{fig:mcdens} displays examples of  
$\rho_\mu$ as a function of $E_0$ 
for different muon thresholds, core distances, and primary masses.
The selection cuts have been applied as to the measured data.
The error bars indicate the width of the distributions. 
They decrease with increasing energy and mass. 
A power law dependence is fitted in a restricted 
energy range to reduce 
the influence of showers with primary energies outside the 
simulation range. \\   
Figure~\ref{fig:deltas} shows the resulting power law 
indices $\delta$ for all core distances 
\begin{figure}[ht]
\hc{\epsfig{file=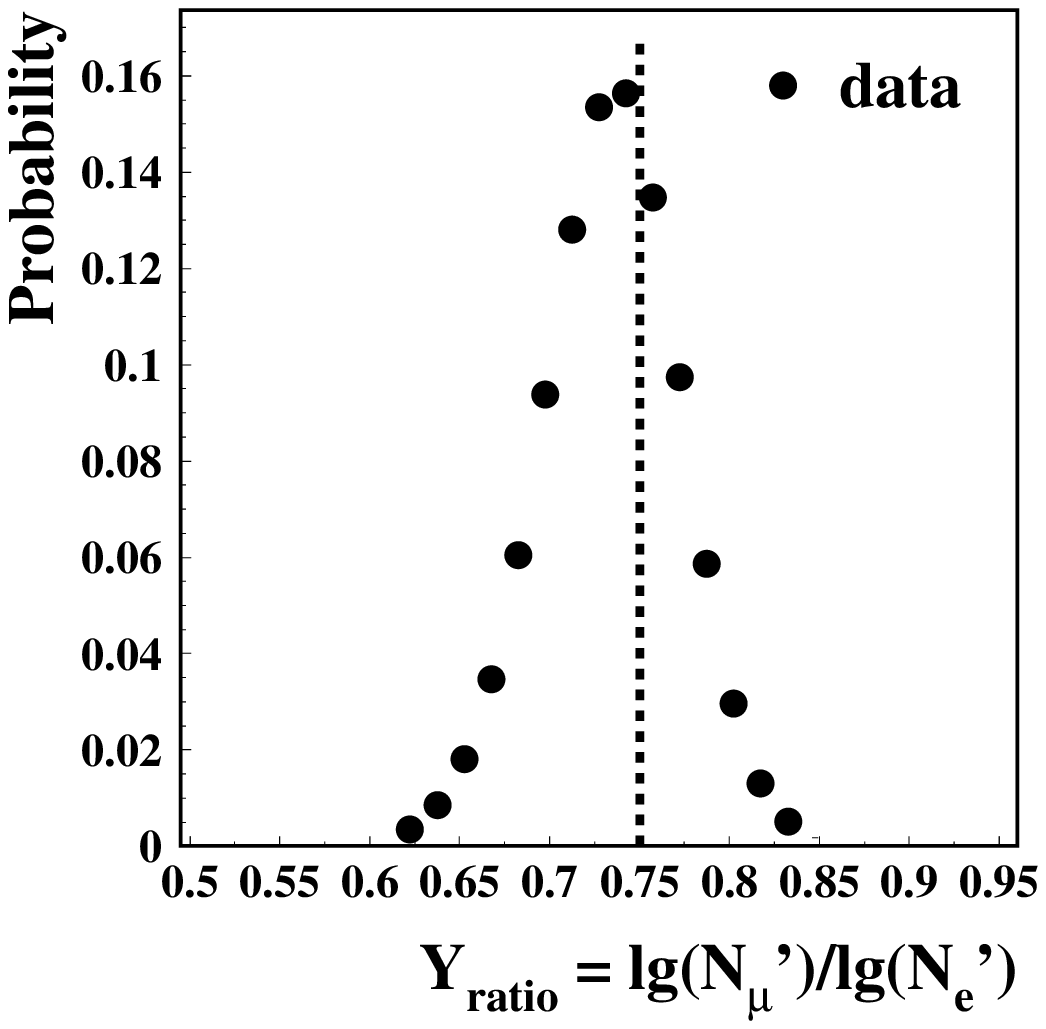,clip=,width=0.5\hsize}
    \epsfig{file=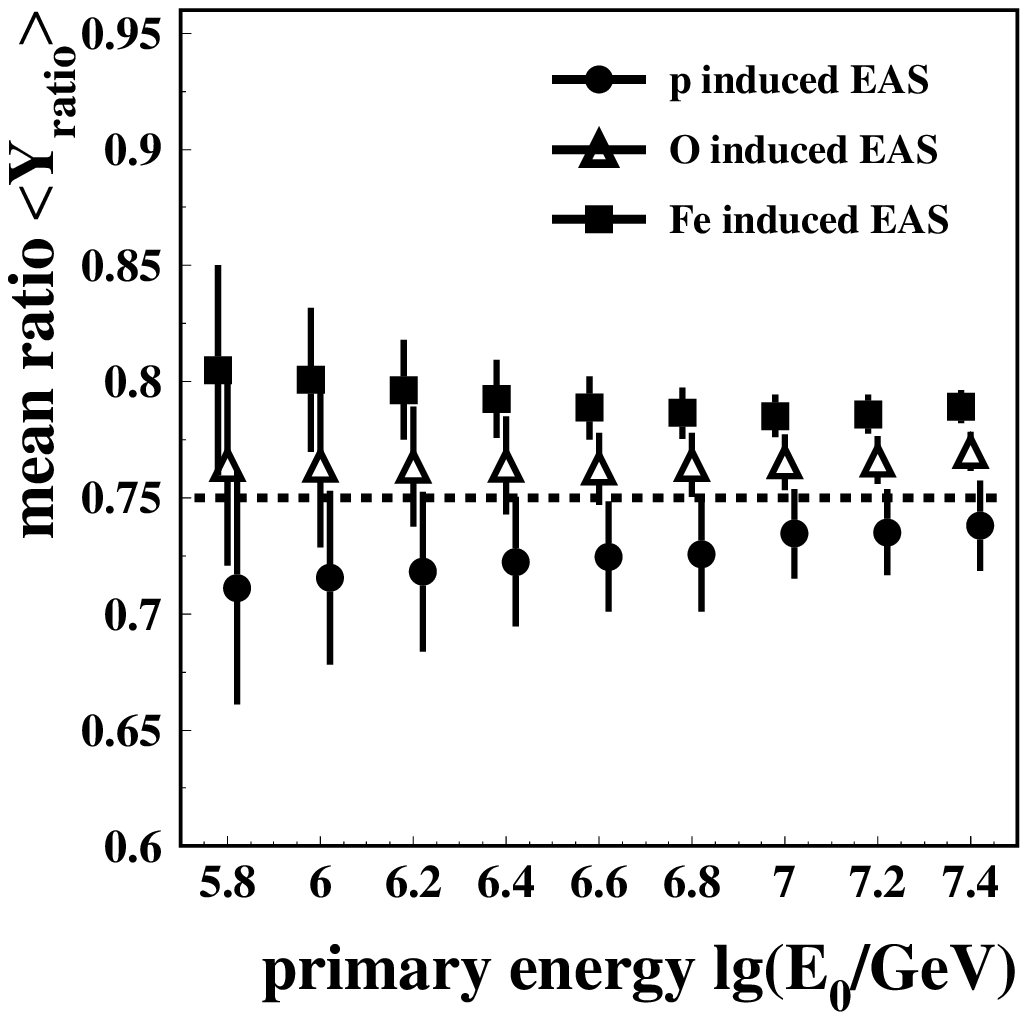,clip=,width=0.5\hsize}}
\caption[EAS separation by particle number ratio]{Distributions
of the parameter $Y_{\rm ratio}$ used for mass separation 
shown for measured (left hand) and simulated (right hand) EAS. 
In case of simulations the energy dependence of the ratio is shown. 
The error bars represent the r.m.s. of the distributions.
The dashed lines indicate the dividing line between 
``electron-rich'' and ``electron-poor''.
All selection cuts are applied.}
\label{fig:ratio}
\end{figure}
in case of primary protons and iron nuclei.
The slopes are nearly independent of the radial distance,
confirming the weak dependence of the shape of the muon lateral
distribution with primary energy~\cite{antoni-lat}. 
But the slopes depend on muon threshold and primary mass. 
Systematic uncertainties are at the 10\% 
level and indicated in Figure~\ref{fig:deltas}. 
These systematics are estimated by varying the energy and angular 
spectrum of the simulated events, as well as modifying the degree
of fluctuation of the observables resulting from the simulations. 
The uncertainty of the high-energy 
interaction model itself is tested with a set of simulations based
on the VENUS high-energy interaction model~\cite{VENUS}.
The slope of the $\rho_\mu \propto E_0^\delta$-dependence 
differs systematically by $\approx +0.1$ for both 
energy thresholds, but this systematics is not included
in the error bars in Figure~\ref{fig:deltas}. \\
The detailed simulations allow also to verify that the 
cut on the shower size ratio 
$Y_{\rm ratio} = lg(N_\mu^\prime) / lg(N_e^\prime)$ 
is energy independent.
The distribution of $Y_{\rm ratio}$ for the
measured events, and the quality of the mass 
separation as provided by the simulation calculations
are shown in Figure~\ref{fig:ratio}. To divide
the total sample in electron-rich and electron-poor EAS 
a cut of $Y_{\rm ratio}=0.75\,$is chosen. 
By this most of the proton induced showers belong to the 
``electron-rich'' class, whereas primary 
iron and medium nuclei are associated to the ``electron-poor'' 
class. Obviously the classification is nearly energy 
independent (Figure~\ref{fig:ratio} right).
Also, the qualitative behavior of the electron-poor and 
electron-rich distributions was found to be insensitive
to small changes of $Y_{\rm ratio}\,$.

\subsection {Features of the energy spectra}
\label{energ}
When relating the density spectra to the primary energy spectrum 
of cosmic rays a power law spectrum 
$\frac{dN}{dE_0} \propto E_0^{-\gamma}$ is assumed. 
The energy spectrum can be written as 
$\frac{dN}{d\rho_\mu} \cdot \frac{d\rho_\mu}{dE_0}$, 
where $\frac{d\rho_\mu}{dE_0}$ has to be deduced from the EAS 
simulations and $\frac{dN}{d\rho_\mu} \propto (\rho_\mu)^{-\beta}$
is taken from the experimental results. 
Thus the spectral index $\gamma$ can be expressed by 
$\gamma = \delta\cdot(\beta-1)+1$  with $\delta$  
from the simulations (${\rho_\mu} \propto E_0^{\,\, \delta}\,$,
see Figure~\ref{fig:deltas}). 
If the correct elemental composition is adopted, all 
measured muon density spectra (of the total sample or
of a certain subsample) should result consistently in the true primary 
energy spectrum, irrespective which core distance and muon 
energy threshold are considered. 
Hence by use of the results of various core distance ranges and 
different muon energy thresholds systematic effects induced by 
the Monte Carlo simulations could be checked. E. g., a possible 
\begin{figure}[ht]
\hc{\epsfig{file=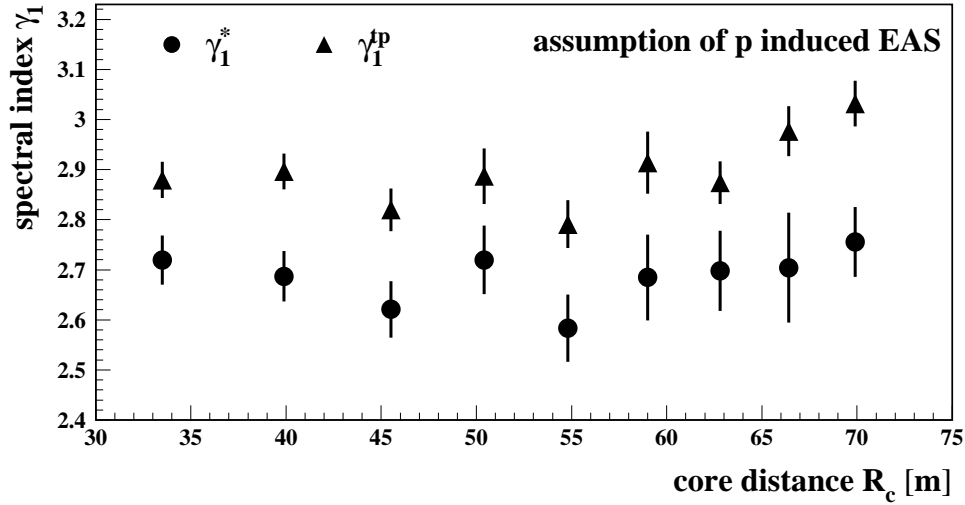,clip=,width=\hsize}}
\caption[Indices of primary energy spectrum]{Variation of the 
reconstructed power law index below the knee of the primary 
all-particle energy spectrum with the core distance for both 
energy thresholds assuming a pure proton composition.}
\label{fig:gammas}
\end{figure}
\begin{figure}[ht]
\hc{\epsfig{file=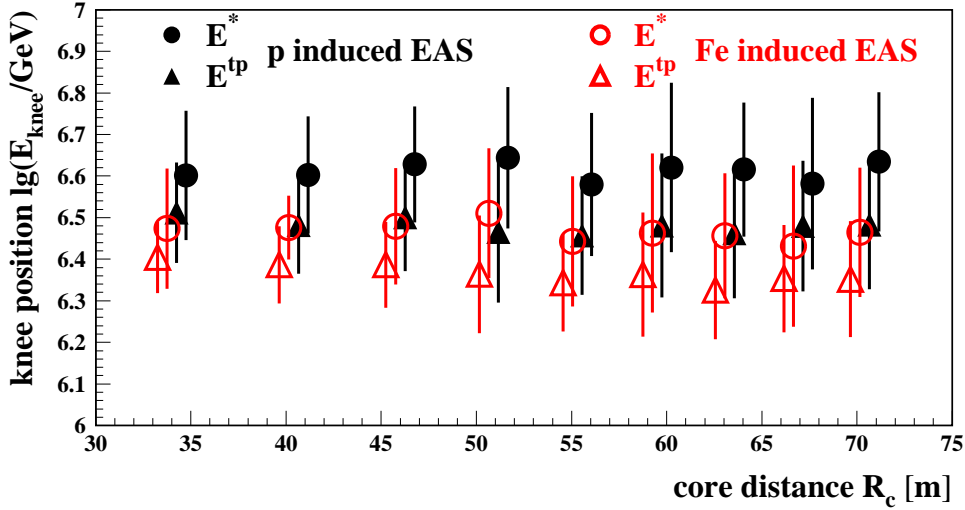,clip=,width=\hsize}}
\caption[Position of the Knee]{Variation of the estimated knee 
position in the primary all-particle spectrum with the 
core distance for both energy thresholds, with the 
assumption of a pure proton or pure iron cosmic ray beam.}
\label{fig:eknee}
\end{figure}
dependence of the slopes and of the knee position on 
core distance would indicate an insufficient 
description of the muon lateral distribution by the simulations 
while a comparison of the spectra observed with different 
muon energy thresholds could reveal inconsistencies of the 
simulated muon energy spectrum. \\
As an example, Figure~\ref{fig:gammas} shows the variation of the 
resulting exponents $\gamma_1$ of the primary all-particle 
spectrum below the knee derived for both muon energy
thresholds under the assumption that the primaries are protons. 
The smaller fluctuations in $\rho_\mu$ with 
increasing atomic number result in a decrease of the spectral
index of the resulting primary energy spectrum for the assumption 
of a heavier composition. 
Figure~\ref{fig:eknee} compares the knee positions determined 
under the assumptions of proton primaries with the results found 
for the case assuming iron primaries. 
The knee positions resulting from iron nuclei as primaries are 
systematically shifted to smaller energies as compared to a 
pure proton composition. This is due to the fact that the 
\begin{table}
\caption[The Results]{The spectral indices, fluxes and
positions of the knee in the primary energy spectrum 
for the different EAS samples analyzed assuming single
element primaries ($dJ_{\rm knee}/dE$ is given in 
$[$m$^{-2}$s$^{-1}$sr$^{-1}$GeV$^{-1}$$]$). The parameters
do not depend on the core distance. Therefore mean values are given.}
\label{tab:result}
\begin{center}
\vspace*{0.3cm}
\begin{tabular}{|c|c|c|c|}
\hline
\multicolumn{4}{|c|}{\rule[-2mm]{0mm}{6mm} all-particle spectrum} \\
\hline
      & {for primary proton} & {for primary oxygen} & {for primary iron} \\
\hline
 { $\gamma_{1}^\star$} 
      & { 2.68 $\pm 0.02$ }   
      & { 2.66 $\pm 0.02$ }
      & { 2.51 $\pm 0.02$ } \\
 { $\gamma_{2}^\star$} 
      & { 2.89 $\pm 0.02$ } 
      & { 2.85 $\pm 0.01$ } 
      & { 2.71 $\pm 0.01$ } \\
 { lg$(E_{\rm knee}^\star/$GeV)}
      & { 6.61 $\pm 0.05$}
      & { 6.53 $\pm 0.05$}
      & { 6.47 $\pm 0.05$}
      \\
 { $dJ_{\rm knee}^\star/dE$}
      & { (10.4 $\pm 2.0$) $\cdot 10^{-14}$}
      & { (12.5 $\pm 2.5$) $\cdot 10^{-14}$}
      & { (13.9 $\pm 2.9$) $\cdot 10^{-14}$}
      \\
\hline                                       
 { $\gamma_{1}^{\rm tp}$} 
      & { 2.89 $\pm 0.01$ } 
      & { 2.96 $\pm 0.01$ } 
      & { 2.75 $\pm 0.01$ } \\
 { $\gamma_{2}^{\rm tp}$} 
      & { 3.11 $\pm 0.02$ } 
      & { 3.18 $\pm 0.01$ } 
      & { 2.96 $\pm 0.01$ } \\
 { lg$(E_{\rm knee}^{\rm tp}/$GeV)} 
      & { 6.48 $\pm 0.05$}
      & { 6.43 $\pm 0.04$}
      & { 6.37 $\pm 0.04$}
      \\
 {$dJ_{\rm knee}^{\rm tp}/dE$ }
      & { (3.0 $\pm 0.8$) $\cdot 10^{-14}$}
      & { (2.3 $\pm 0.3$) $\cdot 10^{-14}$}
      & { (3.7 $\pm 1.1$) $\cdot 10^{-14}$}
      \\
\hline  
\end{tabular}
\vspace*{0.2cm}
\begin{tabular}{|c|c|c|}
\hline
      & electron-rich sample & electron-poor sample \\
\hline
      & {for primary proton}  & {for primary iron} \\
\hline
 { $\gamma_{1}^\star$} 
      & { 2.83 $\pm 0.03$ }   
      & { 2.40 $\pm 0.01$ } \\
 { $\gamma_{2}^\star$} 
      & { 3.41 $\pm 0.03$ } 
      &  \\
 { lg$(E_{\rm knee}^\star/$GeV)}
      & { 6.70 $\pm 0.05$}
      & 
      \\
\hline                                       
 { $\gamma_{1}^{\rm tp}$} 
      & { 2.97 $\pm 0.02$ } 
      & { 2.69 $\pm 0.01$ } \\
 { $\gamma_{2}^{\rm tp}$} 
      & { 3.53 $\pm 0.02$ } 
      &  \\
 { lg$(E_{\rm knee}^{\rm tp}/$GeV)} 
      & { 6.53 $\pm 0.04$}
      & 
      \\
\hline  
\end{tabular}
\end{center}
\vspace*{0.2cm}
\end{table}
local muon density is increasing with the primary mass 
(see Figure~\ref{fig:mcdens} left). \\
The density spectra for the different
core distances are independent of each other 
and the resulting slopes and knee positions of the primary 
energy spectrum agree within their statistical uncertainties. 
This supports the confidence in the lateral distribution 
predicted by the Monte Carlo simulations, and allows
to present results averaged over all core distance bins 
(Table~\ref{tab:result}). 
Nevertheless there remain obvious systematic differences in 
the results for the two muon energy thresholds, observed for 
all core distances. 
The systematic differences might arise from possibly incorrect 
assumptions on mass composition due to the sensitivity of the muon 
spectrum to primary mass. 
Such an effect, however, should be considerably reduced when 
analysing the electron-rich and electron-poor subsamples which 
should be enriched in light and heavy primaries, respectively. 
For these samples no variation with core distance is again 
observed, and Table~\ref{tab:result} presents 
average values. But the systematic differences for the two 
thresholds remain. 
In order to check the influence of heavier contributions 
the electron-rich~EAS sample has been additionally analyzed 
assuming larger fractions of helium and even oxygen 
primaries. A flattening of the spectra and a shift of the knee 
position to lower energies occurs by up to 5\% (with the extreme 
assumption of 100\% oxygen). When assuming a pure oxygen 
composition for the electron-poor sample the primary spectrum 
steepens by $\Delta \gamma \approx 0.15$. Such effects do not 
explain the systematic discrepancy displayed by the results 
from the two different muon energy thresholds. 
Therefore we conclude that an incorrect description of the 
muon energy spectrum by the Monte Carlo simulations is the 
origin of the discrepancy. \\
The effect does not only occur for the QGSJet model used for 
the present analysis. A smaller sample of reference showers 
generated with the VENUS model has been used to study the 
observed difference. A general shift to a steeper primary 
energy spectrum ($\Delta \gamma \approx 0.2$) and a 
lower knee position is found. 
That may be associated to differences in the modelling of 
the high-energy interaction~\cite{antoni-non}. 
However, the inconsistency with respect to the two different 
muon energy thresholds persists. 
The considered muon energies are comparatively low, and are 
treated in the CORSIKA simulation 
code mainly by the low-energy interaction model GHEISHA. 
Thus the inconsistencies are most probably due to the 
low-energy model. 
There are in fact indications for deficiencies of the code 
from another study~\cite{wentz-SLC}.


\section {Summary and conclusions}
\label{sec:concl}

Frequency spectra of local muon densities of EAS
in the PeV region were measured and analyzed for various
core distances and for two muon energy thresholds. 
For both thresholds the all-particle spectra show the knee 
structure, i.e. two power laws with increasing steepness
in the knee region. 
Compared to shower size spectra
based on the charged particle or electron 
number~\cite{eastop-1999,ralph-spec}, the 
muon density spectra show a relatively smooth knee with
a small, but clear change of the power law exponent. \\ 
With help of the muon to electron number ratio, estimated 
on an event-by-event basis, the registered EAS are divided into  
electron-rich and electron-poor subsamples. 
The subsample of the electron-rich EAS shows the same knee features 
as the total sample but with a more pronounced knee. 
The electron-poor sample shows no change
of slope within the density range investigated.
An identical feature has also been observed 
in combined energy and composition analyses of  
size spectra measured by KASCADE~\cite{kampert-SLC} albeit then
based on Monte Carlo simulations. 
Simulations indicate that electron-rich showers originate 
from light primary nuclei.
This feature holds irrespectively of details of the 
interaction models.
Hence we conclude that the knee reflects a feature of the light
particle spectrum and that the spectrum of heavy particles
does not change slope in the range of our measurements.
Such a behaviour is expected if the knee
is caused either by interstellar magnetic fields or a change of the
interaction in the atmosphere since in these cases the knee of nuclei
should be displaced by a factor of Z or A, respectively, to that
of protons. \\
Detailed EAS and detector simulations were used to interpret
the measured muon density spectra in terms of the primary 
energy spectrum.
Independent of the elemental composition assumed all measured spectra
should result in the same primary energy spectrum, irrespectively of
core distance or muon threshold. 
However, only when assuming the true composition the derived energy
spectrum will be the correct one.
This agreement for different spectra actually is observed 
for varying core distances.
Hence we conclude that the muon lateral distribution 
is sufficiently well represented by the simulations.  
In contrast, the results for the two energy thresholds lead
to different exponents and positions of the kinks.
These differences are larger than the systematic uncertainties
due to the unknown composition, especially in the case of
the subsamples.
Thus, the measurements presented here reveal that the Monte Carlo
simulations are not capable to describe the muon energy distribution
correctly.
Such indications arise also from studies of the muon lateral
distributions for different muon energy thresholds with
KASCADE~\cite{haungs-fzka}. 
In view of these systematic discrepancies, 
it is difficult to draw definite conclusions 
but some general features of 
the primary energy spectrum can be stated:
The all-particle energy spectrum exhibits a knee 
at $E_{\rm knee} \approx (3-5) \cdot 10^{15}\,$eV
with a change of the spectral index of order 
$\Delta \gamma \approx 0.2-0.3\,$.
This knee is only seen in the light ion subsample, at the same 
position but with a distinctly larger steepening of 
$\Delta \gamma \approx 0.5\,$.
The heavy ion component of the cosmic ray flux displays no 
steepening in the energy range of 1--10 PeV and a smaller slope
than the light component below the knee. 
Within the uncertainties the findings about the all-particle
spectrum are compatible with recent results from 
KASCADE~\cite{antoni-non} and other 
experiments~\cite{casa-1999,tibet-1993,akeno-1984}. \\
From the experimental point of view the study of muon density 
spectra establishes a new approach to investigate the energy 
spectrum of cosmic rays.
Although the statistical accuracy is limited 
the measured spectra reflect the features 
of the primary energy spectrum in an astonishingly direct manner. 
But considerable inconsistencies arise when attempting to convert
the measured muon density spectra into the primary energy spectrum
based on simulations. 
It is only due to multiparameter measurements of experiments
such as KASCADE that the deficiencies of the simulations get 
revealed. 
Thus improvements of the hadronic interaction models 
incorporated in the simulations appear to 
be the most important prerequisite for a consistent 
interpretation of the data in 
terms of elemental composition and energy spectrum of primary
cosmic rays in the knee region.    

\section*{Acknowledgments}

The authors would like to thank the 
members of the engineering and technical staff of the KASCADE 
collaboration who contributed to the 
success of the experiment with enthusiasm and commitment.
The work has been supported by the Ministery for Research of the 
Federal Government of Germany, by a grant of the 
Romanian National Agency for Science, Research and Technology as 
well as by a research grant (No. 94964) of the Armenian Government 
and by the ISTC project A116. The collaborating group of the 
Cosmic Ray Division of the Soltan Institute of Nuclear Studies 
in Lodz and of the University of Lodz is supported by the Polish 
State Committee for Scientific Research. The KASCADE collaboration 
work is embedded in the frame of scientific-technical cooperation 
(WTZ) projects between Germany and Armenia (No. 002-98),
Poland (No.POL-99/005), and Romania (No.RUM-014-97).


\ifx\undefined\JournalLength\def\JournalLength{M}\fi
  \def\NewJournal#1#2#3#4{\ifx\undefined#1 \ifx S\JournalLength
  \newcommand{#1}{#2~} \else \ifx L\JournalLength \newcommand{#1}{#4~} \else
  \newcommand{#1}{#3~}\fi\fi\fi} \NewJournal{\araa}{Ann. Rev. A\&A}{Ann. Rev.
  Astr. Astrophys.}{Annual Review of Astronomy and Astrophysics}
  \NewJournal{\aj}{AJ}{Astronom. J.}{Astronomical Journal}
  \NewJournal{\aap}{A\&A}{Astron. Astrophys.}{Astronomy and Astrophysics}
  \NewJournal{\aapr}{A\&A. Rev.}{Astron. Astrophys. Rev.}{Astronomy and
  Astrophysics Reviews} \NewJournal{\apj}{ApJ}{Astrophys. J.}{Astrophysical
  Journal} \NewJournal{\apjl}{ApJ Lett.}{Astrophys. J. Lett.}{Astrophysical
  Journal, Letters to the Editor} \NewJournal{\apjs}{ApJ Suppl.}{Astrophys. J.
  Suppl.}{Astrophysical Journal, Supplement Series}
  \NewJournal{\apss}{ApSS}{Astroph. Space Sci.}{Astrophysics and Space Science}
  \NewJournal{\baas}{BAAS}{Bull. American Astron. Soc.}{Bulletin of the
  American Astronomical Society} \NewJournal{\iaucirc}{IAUC}{IAU Circ.}{IAU
  Circular} 
  \NewJournal{\jphysg}{J. Phys. G}{J. Phys. G: Nucl. Part. Phys.}
  {Journal of Physics G: Nuclear and Particle Physics}
  \NewJournal{\mnras}{M.N.R.A.S.}{Mon. Not. Royal Astron. Soc.}{Monthly Notices
  of the Royal Astronomical Society} \NewJournal{\nat}{Nature}{Nature}{Nature}
  \NewJournal{\nim}{NIM}{Nucl. Instr. and Meth.}{Nuclear Instruments and
  Methods} \NewJournal{\nphysa}{Nucl. Phys. A}{Nucl. Phys. A}{Nuclear Physics
  A} \NewJournal{\nphysb}{Nucl. Phys. B}{Nucl. Phys. B}{Nuclear Physics B}
  \NewJournal{\pra}{Phys. Rev. A}{Phys. Rev. A}{Physical Review A}
  \NewJournal{\prb}{Phys. Rev. B}{Phys. Rev. B}{Physical Review B}
  \NewJournal{\prc}{Phys. Rev. C}{Phys. Rev. C}{Physical Review C}
  \NewJournal{\prd}{Phys. Rev. D}{Phys. Rev. D}{Physical Review D}
  \NewJournal{\pre}{Phys. Rev. E}{Phys. Rev. E}{Physical Review E}
  \NewJournal{\prl}{Phys. Rev. Lett}{Phys. Rev. Lrtt.}{Physical Review Letters}
  \NewJournal{\physrep}{Phys. Rep.}{Phys. Rep.}{Physics Reports}
  \NewJournal{\skytel}{Sky \& Tel.}{Sky \& Tel.}{Sky and Telescope}
  \ifx\undefined\solar\def\solar{\hbox{$_{{\scriptscriptstyle \odot}}$}}\fi
  \ifx\undefined\arcseconds\def\arcseconds{\makebox[0pt]{\hskip
  0.4em$^{\prime\prime}$}}\fi

\end{document}